\documentstyle[prb,aps,multicol,epsfig,float]{revtex}
\newcommand{\nirt}{n_i(\vec{r},t)}
\newcommand{\nieq}{n_{i}^{eq}(\rho,\vec{u})}
\renewcommand{\vec}[1]{{\mathbf{#1}}}
\begin{document}
\draft

\title{Simulation of a Single Polymer Chain in Solution by
Combining Lattice Boltzmann and Molecular Dynamics}
\author{Patrick Ahlrichs, Burkhard D\"unweg}
\address{Max Planck Institute for Polymer Research, Ackermannweg 10, D-55128
  Mainz, Germany}

\date{\today}

\maketitle

\widetext
\begin{abstract}
  In this paper we establish a new efficient method for simulating
  polymer--solvent systems which combines a lattice Boltzmann approach
  for the fluid with a continuum molecular dynamics (MD) model for the
  polymer chain. The two parts are coupled by a simple dissipative
  force while the system is driven by stochastic forces added to both
  the fluid and the polymer. Extensive tests of the new method for
  the case of a single polymer chain in a solvent are performed. The
  dynamic and static scaling properties predicted by analytical theory
  are validated. In this context, the influence of the finite size of
  the simulation box is discussed. While usually the finite size
  corrections scale as $L^{-1}$ ($L$ denoting the linear dimension of
  the box), the decay rate of the Rouse modes is only subject to an
  $L^{-3}$ finite size effect. Furthermore, the mapping to an existing
  MD simulation of the same system is done so that all physical input
  values for the new method can be derived from pure MD
  simulation. Both methods can thus be compared quantitatively,
  showing that the new method allows for much larger time
  steps. Comparison of the results for both methods indicates
  systematic deviations due to non--perfect match of the static chain
  conformations.
\end{abstract}


\begin{multicols}{2}
\narrowtext

\section{Introduction}

The complexity and variety of soft condensed matter is largely due to
the fact that length scales of different orders of magnitude are
present \cite{degennes:79,strobl:97}. When dealing with polymers in
computer simulations, one therefore often intends to analyze the
scaling behavior, where the nature of the underlying chemistry becomes
unimportant \cite{degennes:79,doi:86}. When constructing models for
these systems it is crucial to coarse--grain the details and to keep
the relevant length scales in order to observe the phenomena one is
interested in. Since bead--spring models in MD simulations are an
appropriate means to yield the right universal laws, they have been
widely used to simulate the scaling behavior of polymers and much
progress has been made using these models
\cite{binder:95,minn:98,kremer:90,duenweg:93,pierleoni:92,smith:92}.

While in some systems, e.~g. in highly concentrated solutions or in
melts, the dynamic properties are not affected by the solvent --- such
that these can be simulated by conventional bead--spring models
without explicitly taking into account the solvent --- there are many
phenomena in polymer science where the influence of the solvent on the
polymer dynamics cannot be neglected. For example, in dilute or
semi--dilute polymer solutions, the dynamical behavior is changed and
even dominated by hydrodynamic interaction between different parts of
the polymers. This eventually leads to a long--range interaction which
is mediated by the solvent \cite{degennes:79,doi:86}. With this paper,
we want to provide a new efficient method for the simulation of
polymer systems where hydrodynamics plays a role. The idea is to focus
on the really necessary parts only, i.~e. the hydrodynamics of the
solvent and the (Brownian) motion of the polymer chains, thereby
trying to keep the computational costs at a minimum.  Our test case is
the dynamics of a single chain in a solvent. This problem has
continuously attracted the attention of MD researchers
\cite{duenweg:93,pierleoni:92,smith:92}, mainly because existing
analytical theories \cite{kirkwood:48,kirkwood:58,zimm:56} rely on
uncontrollable assumptions that can be tested using computer
simulations.

Simulating such systems by MD is only possible if one introduces
explicit solvent particles. Hence one has to face the problem that
almost all CPU time goes into the propagation of the solvent
particles, while one is mainly interested in the chain properties.
However, there are also other computational methods than MD available
for soft condensed matter systems where hydrodynamics is important,
not only in the field of polymers but for example also in colloidal
suspensions. These include Brownian Dynamics simulations
\cite{ermak:78,brady:88,oettinger:89,rey:91}, and Dissipative
Particle Dynamics (DPD) \cite{hoogerbrugge:92,schlijper:95,%
warren:97,espanol:95,marsh:97,espanol:98,ignacio:98}. Both of them
have inherent strengths, but also some disadvantages: The first
technique must face the problem that the algorithm scales as the cube
of the number of particles, and the latter (like MD) simulates the
solvent particles explicitly, leading to simulations of several
thousand particles even for a single chain of, say, 30 monomers.
Compared to MD, DPD has the advantage of much larger time steps,
mainly because of the use of very soft potentials \cite{warren:97}. A
lot of progress in the theoretical framework of the method has been
achieved \cite{espanol:95,marsh:97,espanol:98}, but some practical
problems remain, like the time step dependent temperature and the
small Schmidt number \cite{warren:97}. Recently, however, some effort
has been made to fill this gap \cite{ignacio:98}.

In this paper we use a recently proposed method \cite{ahlrichs:98}
that couples a lattice Boltzmann approach for the fluid to
bead--spring polymer chains. The lattice Boltzmann method (LBM)
\cite{benzi:92,chen:98} was developed to simulate hydrodynamics on a
grid. The LBM was shown to be an effective and fast method for
simulating fluid flows, comparable to finite--difference
\cite{ladd2:94} or spectral methods \cite{martinez:93}. Ladd applied
the LBM successfully to colloidal systems \cite{ladd2:94,ladd:94}: The
colloidal particles are simulated as hard spheres by using stick
boundary conditions. This leads to a very efficient algorithm: Its CPU
cost scales linearly with the number of particles, and it uses a
``minimal'' model to simulate the fluid. Besides, Ladd also
showed \cite{ladd:94} that fluctuations can be incorporated into the
LBM in the spirit of Landau--Lifshitz fluctuating
hydrodynamics \cite{landau:59}, which is essential if one wants to
investigate Brownian motion.

Now one might think of a direct application of Ladd's method to
polymer--solvent systems. However, using hard spheres to model the
monomers is not necessary here, as rotational degrees of freedom as
well as stick boundary conditions are not relevant: On the large
length and time scales we are interested in, like the radius of
gyration and the Zimm time of the polymer, it is sufficient that
hydrodynamic interaction has evolved. The ``microscopic details'' of
the coupling should then not play a role. In this spirit, we couple
the LBM to bead--spring polymer chains by a simple friction ansatz,
thereby treating the monomers as point particles for the fluid. We
will show that this ansatz is sufficient to simulate both the static
and dynamic scaling behavior of the polymer. The simulation of the
fluid by LBM rather than explicit particles and the simple friction
ansatz lead to a large speedup in computer time of about a factor of
20 when compared to pure MD, or even more if one is willing to be
satisfied with less accurate data.
   
Additionally, we map our method to a pure MD simulation, i.~e. we show
how to determine all physical input values from the results of MD,
allowing us to compare our results to an existing MD simulation with
explicit solvent particles \cite{duenweg:93}. In other words, the
fluid in the new method can be viewed as a coarse--grained MD fluid,
and there exists a well--defined procedure for how to do the
coarse--graining. Of course, in using such a mesoscopic approach it is
no longer possible to include detailed chemistry like in atomistic MD
simulations. This is, however, a quite common feature of mesoscopic
simulation methods; DPD simulations do not include atomistic details
either.

The remainder of this article is organized as follows: We outline the
method in Section \ref{method}, and present the numerical results in
Section \ref{sec:singlechain}, which are compared to pure MD in
Section \ref{sec:mapping}. In Section \ref{sec:conclusion} we conclude
with some final remarks and an outlook to further studies.

\section{The Simulation Method}\label{method}

\subsection{The Lattice Boltzmann Method for the Solvent}

The lattice Boltzmann method is a discrete formulation of the
Boltzmann equation on a lattice, leading to the Navier--Stokes
equations in the incompressible limit by means of a Chapman--Enskog
expansion \cite{benzi:92,chen:98}. It has been successfully applied to
a variety of fluid flow problems, and it is especially well--suited
for complex fluids because of the possibility of straightforward
implementation of complex boundaries. The central quantity of the
algorithm is $\nirt$, the number of particles in a volume $a^{3}$ at
the grid point $\vec{r}$ at time $t$, which have the velocity
$\vec{c}_{i}\frac{a}{\tau}\,\, (i=1,..,b)$, where $a$ is the lattice
spacing, $\tau$ the time step and $\vec{c}_{i}$ a vector leading to
the $i$th neighbor on a grid with unit lattice constant. The evolution
equation for $\nirt$ is the lattice Boltzmann equation
\begin{eqnarray}
n_{i}(\vec{r}+\vec{c}_{i}a, t+\tau)=&&\nirt\\\nonumber
&&+\sum_{j=1}^{b}
L_{ij}\left(n_{j}(\vec{r},t)-n_{j}^{eq}(\rho,\vec{u})\right).
\end{eqnarray}   
The last term expresses the relaxation of $n_{i}$ towards a local
pseudo--equilibrium, which resembles a Bhatnagar--Gross--Krook (BGK)
collision operator \cite{bhatnagar:54} in the continuum Boltzmann
equation. The constant matrix $L_{ij}$ can be interpreted as the
scattering between particle population $i$ and $j$. Its eigenvalues
can be determined from physical and numerical arguments, such that its
explicit form is not necessary for the simulation algorithm
\cite{ladd:94}. The local pseudo--equilibrium dis\-tri\-bution $\nieq$
de\-pends on the density $\rho(\vec{r},t)=\sum_i\nirt\mu/a^{3}$ and
fluid current $\vec{j}(\vec{r},t)\equiv\rho\vec{u}=\sum_i\nirt
\vec{c}_{i}\mu/(\tau a^{2})$ only. Here, $\mu$ is the mass of a fluid
particle. The usual functional form for $\nieq$ is assumed
\cite{chen:98}:
\begin{equation}\label{eq:eq.distr.}
\nieq =
\rho\left( A_{q}+B_{q}\left( \vec{c}_{i}\cdot\vec{u}\right)
  +C_{q}u^{2}+D_{q}\left( \vec{c}_{i}\cdot\vec{u}\right)^{2} \right).
\end{equation}
The coefficients $A_{q}$, $B_{q}$, $C_{q}$ and $D_{q}$ (which depend
on the sublattice $q$, i.~e. the magnitude of $\vec{c}_{i}$) are
determined to reproduce the correct macroscopic hydrodynamic
behavior. Note that this is contrary to continuum kinetic theory,
where the Maxwell--Boltzmann distribution is determined by entropy
considerations and the Navier--Stokes equations follow naturally by
the Chapman--Enskog expansion \cite{hirschfelder:64,wagner:98}. Hence
it is called {\em pseudo}--equilibrium. Explicit values for the
coefficients $A_{q}$,$B_{q}$,$C_{q}$ and $D_{q}$ are known for
different lattices \cite{qian:92}.

Here, we implement the 18--velocity model of Ref.
\onlinecite{ladd:94}, which corresponds to the D3Q18 model in the
nomenclature of Ref. \onlinecite{qian:92}. The set of $\vec{c}_{i}$
consists of the 6 nearest and 12 next--nearest neighbors on a simple
cubic lattice. Via a Chapman--Enskog expansion one can show that this
model leads to the Navier--Stokes equations in the limit of small
Knudsen and Mach numbers \cite{benzi:92}, and derive a relation
between the kinematic viscosity $\nu$ and the non--trivial eigenvalue
$\lambda$ of $L_{ij}$ belonging to the eigenvector
$c_{i\alpha}c_{i\beta},\,\, (\alpha, \beta = x,y,z,
\alpha\neq\beta)$ \cite{qian:92},
\begin{equation} \label{eq:relationnulambda}
\nu = -\frac{1}{6}\left(\frac{2}{\lambda}+1\right)\frac{a^{2}}{\tau}.
\end{equation}

In this paper, we always deal with low Reynolds number flow, hence the
linearized Navier--Stokes equations are sufficient. For this reason,
we neglect the nonlinear term in the equilibrium distribution
(\ref{eq:eq.distr.}), i.~e. we effectively set $C_{q} = D_{q} = 0$,
thus obtaining a simpler and faster algorithm \cite{ladd:94}.

Fluctuations can be incorporated into the lattice Boltzmann method
\cite{ladd:94}. The central idea is to add fluctuations to the fluxes
of the conserved variables, i.~e. the stress tensor, and not to the
hydrodynamic fields $\rho$ and $\vec{j}$. In this way, local mass and
momentum conservation can be guaranteed \cite{landau:59}. The
fluctuating lattice Boltzmann equation reads
\begin{eqnarray}\label{eq:stoch.LBM}
n_{i}&&(\vec{r}+\vec{c}_{i}a, t+\tau)=\nirt\\\nonumber 
&&+\sum_{j=1}^{b} L_{ij}\left(n_{j}(\vec{r},t)-n_{j}^{eq}(\rho,\vec{u})\right)
+ n_{i}^{\prime}(\vec{r},t)
\end{eqnarray}
with the stochastic term
\begin{equation}
n_{i}^{\prime}(\vec{r},t)=-D_{q}\sum_{\alpha\beta}
\sigma_{\alpha\beta}^{\prime}c_{i\alpha}c_{i\beta}.
\end{equation}
The random stress fluctuations $\sigma_{\alpha\beta}^{\prime}$ are
assumed to have white noise behavior
\begin{eqnarray}
\left\langle\sigma_{\alpha\beta}^{\prime}(\vec{r},t)
\sigma_{\gamma\delta}^{\prime}(\vec{r}^{\prime},t^{\prime})\right\rangle
  = & & A
  \delta_{\vec{r}\vec{r}\,^{\prime}}\delta_{tt^{\prime}} \\ \nonumber
&&\left(\delta_{\alpha\gamma}\delta_{\beta\delta}+
\delta_{\alpha\delta}\delta_{\beta\gamma}-
\frac{2}{3}\delta_{\alpha\beta}\delta_{\gamma\delta}\right).
\end{eqnarray}
By solving the resulting discrete Langevin equation for the current
one finds the fluctuation--dissipation relation \cite{ladd:94} for
this system; the noise strength $A$ is given by
\begin{equation}\label{stoch.latt.boltz.}
A = \frac{2\eta k_{B}T\lambda^{2}}{a^{3}\tau},
\end{equation}
where $\eta\equiv \nu\rho$ is the dynamic viscosity.

The LBM was tested extensively, compared to other Navier--Stokes
solvers and found to have comparable speed and accuracy (see for
example Refs. \onlinecite{benzi:92,ladd2:94,martinez:93,ladd:94}).

\subsection{The Bead--Spring Model for the Polymer Chain}
\label{sec:bead.spring}

The polymer model consists of repulsive Lennard--Jones monomers
connected via non--harmonic springs (FENE potential) \cite{kremer:90}:
\begin{eqnarray}
  \label{LJ}
  V_{\text{LJ}} & = & 4\epsilon\left(\left( {\sigma \over r} \right)^{12}
               - \left( {\sigma \over r} \right)^{6}+ \frac{1}{4}\right)
             \hspace{0.5cm}\,\,\,(r<2^{1/6}\sigma )\\ \nonumber
  V_{\text{FENE}} & = &  - {k R_{0}^{2} \over 2}
             \ln\left(1-\left( {r \over R_{0}} \right)^{2}\right) 
             \hspace{0.5cm}\,\,\,\,\,\,(r<R_{0}).
\end{eqnarray}
In order to model the excluded volume effect the Lennard--Jones
potential acts between all monomers.  As usual, the parameters
$\epsilon$, $\sigma$ and the mass $m$ of the monomer define our unit
system. Therefore we wrote the LBM in dimensional form in the last
section, rather than using the usual dimensionless lattice units. The
equations of motion resulting from these potentials are integrated
using the velocity Verlet algorithm
\cite{frenkel:96} with a time step $\Delta t$. Note that there is a
priori no need to set $\Delta t = \tau$ and we will exploit this fact
below.

The polymer model has been applied successfully to the simulation of
many systems \cite{binder:95,minn:98,kremer:90} including a single
chain in explicit solvent \cite{duenweg:93}, so that we can compare
chain properties in using these potentials.

\subsection{Coupling of Fluid and Monomer}
\label{sec:coupling}

As mentioned above, for the length and time scales of the polymer
chain, the ``microscopic'' details of the coupling should not play a
role, as long as one assures that hydrodynamics evolves in the fluid
on time scales faster than the diffusion time scale of the
monomers. It is not necessary to resolve the shape of the monomer for
the fluid. Thus, we can treat one monomer as a point particle. In
analogy to the Stokes formula for a sphere in a viscous fluid, we
assume the force on the monomer exerted by the fluid to be
proportional to the difference of the velocity of the monomer
$\vec{V}$ and the fluid velocity $\vec{u}$ at the monomer's position,
\begin{equation}\label{eq:ansatz}
\vec{F}_{fl}=-\zeta\left[\vec{V}-\vec{u}(\vec{R},t)\right].
\end{equation}
Here, $\zeta$ is a proportionality coefficient which we will refer to
as the ``bare'' friction coefficient. This ansatz has also been used
in the simulation of sedimentation \cite{herrmann:96}.

\begin{figure}
\begin{center}
\epsfig{file=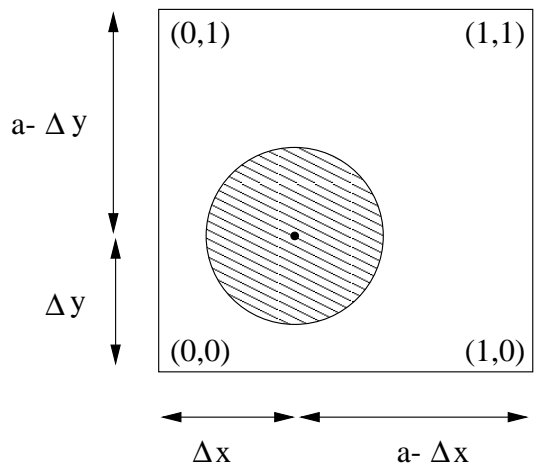,width=7cm}
\end{center}
\caption{Illustration of the quantities used for the coupling of monomer and
  fluid (in two dimensions). The figure shows a sketch of a monomer
  surrounded by the elementary cell of the four nearest neighbor grid
  points. $a$ is the lattice constant.}
\label{fig:coupling}
\end{figure}

Because the fluid velocity is only calculated at the discrete lattice
sites in the simulation, one has to interpolate to get
$\vec{u}(\vec{R},t)$ at the monomer's position. We implement a simple
linear interpolation using the grid points on the elementary lattice
cell containing the monomer: Denoting the relative position of the
monomer in this cell by $(\Delta x,\Delta y, \Delta z)$, with the
origin being at the lower left front edge (see
Fig. \ref{fig:coupling}), we can define
\begin{eqnarray}\label{eq:deltar}
\delta_{(0,0,0)}&=&(1-\Delta x/a)(1-\Delta y/a)(1-\Delta z/a),\\\nonumber
\delta_{(1,0,0)}&=&\Delta x/a\cdot(1-\Delta y/a)(1-\Delta z/a),
\end{eqnarray}
etc. The formula for the linear interpolation then reads
\begin{equation}
\vec{u}(\vec{R},t) = \sum_{\vec{r}\in {\rm ng}} \delta_{\vec{r}}
  \vec{u}(\vec{r},t)
\end{equation}
where ng denotes the grid points on the considered elementary lattice
cell.

In order to conserve the total momentum of fluid and monomer we have
to assign the opposite force to the fluid in that cell. Note that then
the interaction is purely local. In particular, the force density
$-\vec{F}_{fl}/a^{3}$ which is to be given to the fluid leads to a
momentum density transfer per MD time step $\Delta t$ of
\begin{equation}\label{eq:mom.exchange}
-\vec{F}_{fl}/a^{3}=
\frac{\Delta\vec{j}}{\Delta t}=
\sum_{i,\vec{r}\in {\rm ng}}\Delta n_{i}(\vec{r},t)
\vec{c}_{i}\frac{\mu}{a^{2}\tau \Delta t}.
\end{equation}
The last equation has to be satisfied for the change in the number of
particles $\Delta n_{i}$ of the grid points on the elementary lattice
cell in order to exchange the momentum density $\Delta\vec{j}$.
Besides, one must also ensure mass conservation in the fluid,
\begin{equation}
\sum_{i,\vec{r}\in {\rm ng}}\Delta n_{i}(\vec{r},t) = 0.
\end{equation}
The way how to calculate the corresponding $\Delta n_{i}$ at the
nearest grid points is not unique; one possibility was presented in
Ref. \onlinecite{ahlrichs:98}. Here, we follow a different approach
which seems slightly more natural: For given hydrodynamic fields
$\rho(\vec{r},t)$ and $\vec{j}(\vec{r},t)$ at a certain grid point
$\vec{r}$, the equilibrium distribution can be calculated according to
Eq. \ref{eq:eq.distr.}. The change in the equilibrium distribution at
the points $\vec{r}\in{\rm ng}$ due to the presence of the monomer can
therefore be determined: $\rho$ stays constant (mass conservation),
while $\vec{j}\rightarrow\vec{j}+\delta_{\vec{r}}\Delta\vec{j}$. Here
$\delta_{\vec{r}}$ is the fraction~(\ref{eq:deltar}) of the total
$\Delta\vec{j}$ which is given to the specific grid point $\vec{r}$.
Therefore, by requiring that $n_{i} - n_{i}^{eq}$ remains unchanged,
we obtain
\begin{equation}
\Delta n_{i}(\vec{r})=B_{q}\delta_{\vec{r}}\Delta\vec{j}\cdot\vec{c}_{i},
\end{equation}
where again the nonlinear part of Eq. \ref{eq:eq.distr.} has been
neglected, consistent with our overall procedure. More accurate
algorithms (which would however be computationally more expensive)
could be constructed, using the method proposed in Ref.
\onlinecite{martys:98}; however, this is not necessary for our
purposes: Our simple approach is consistent with locality of the
interaction, plus momentum conservation, and should therefore suffice
to build up hydrodynamic interactions in the correct manner.

As we discussed in Ref. \onlinecite{ahlrichs:98}, one has to take care
when adding stochastic terms to the system. Due to the dissipative
nature of the coupling, it is necessary to incorporate fluctuations to
both the fluid and the monomers, i.~e.  to the LBM like in
Eq. \ref{eq:stoch.LBM}, and to the monomers by extending Eq.
\ref{eq:ansatz} to
\begin{equation}\label{eq:ansatz.2}
\vec{F}_{fl}=-\zeta\left[\vec{V}-\vec{u}(\vec{R},t)\right]+\vec{f}.
\end{equation}
Here $\vec{f}$ is a stochastic force of zero mean and 
\begin{equation}
\left\langle{f}_{\alpha}(t){f}_{\beta}(t^{\prime})\right\rangle 
= \delta(t-t^{\prime}) 2 \delta_{\alpha\beta} k_B T \zeta.
\end{equation} 
The momentum transfer to the fluid for the fluctuating case is
calculated in the same way as described above without the
fluctuations. For this reason, the total momentum of fluid and polymer
is conserved locally also in the fluctuating case. One can show
analytically that with this method the fluctuation--dissipation
relation holds for the continuum limit of the model, where the
coupling to the LBM fluid is replaced by the analogous coupling to a
Navier--Stokes fluid with thermal fluctuations of the flow field. For
the velocities of the monomers and the fluid flow velocity, the
equilibrium distribution is then given by the Maxwell--Boltzmann
distribution, while the conformational statistics of the chain is
given by the Boltzmann distribution, i.~e. governed by the
intra--chain potentials $V_{LJ}$ and $V_{FENE}$, see
Eq. \ref{LJ}. This should be contrasted with the MD case, where the
potential due to the solvent particles has an additional
influence. For the discrete case, one can check the
fluctuation--dissipation relation by investigating the velocity
relaxation of one (initially kicked) monomer in the fluid on the one
hand, and the velocity autocorrelation, if fluctuations are added, on
the other hand. The two quantities coincide for our model
\cite{ahlrichs:98}, which is expected from linear response theory. It
is also interesting to note that in the overdamped limit for the
monomer motion, and the continuum limit for the fluid, our approach is
identical to the Oono--Freed equations of motion \cite{oono:81}, which
are commonly used in polymer solution theory.

The main justification of our approach relies on the fact that a
hydrodynamic (Navier--Stokes) description of the fluid works down to
very short (actually, surprisingly short) length and time scales.
Therefore, one should expect that the flow around a monomer should be
describable by the solution of the Navier--Stokes equation as soon as
the distance is larger than a few lattice spacings. The same argument
holds for the analogous MD system, where one expects Navier--Stokes
behavior beyond a few particle diameters. Therefore, we may say that
any two local couplings (for example, our LBM friction ansatz vs. MD)
are equivalent as soon as they produce the same long--range flow
field. If this is the case, then the hydrodynamic interaction between
two monomers (as long as they are not too close) will be identical,
and the single--monomer mobilities will also match (note that for a
particle which is pulled through the fluid at constant velocity by a
constant force, the friction coefficient is determined by the energy
dissipated in the surrounding flow field).

\begin{figure}
\begin{center}
\epsfig{file=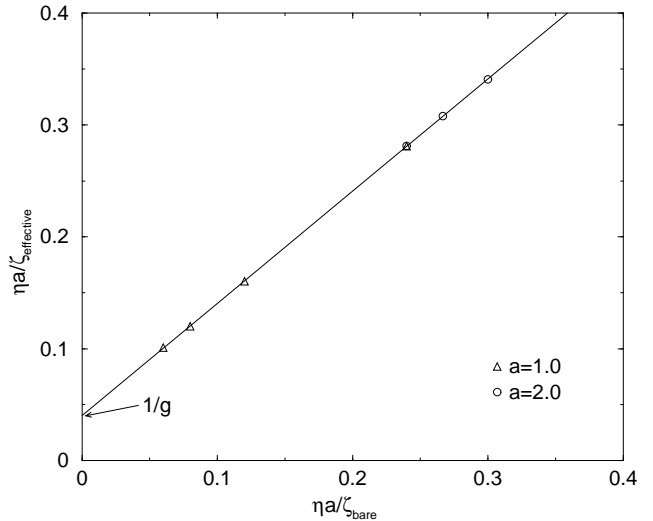,width=7cm,angle=-90}
\end{center}
\caption{Test of the predicted relation between bare and effective
  friction coefficient, Eq. \ref{eq:relationbareeff}. Grids of
  different lattice spacings were used as indicated in the figure.}
\label{fig:zetaeffzetabare}
\end{figure}

This latter property actually allows for an easy determination of the
simulation parameter $\zeta$, which we will now, for the sake of
clarity, denote by the symbol $\zeta_{\rm bare}$. A heuristic
procedure, which was followed in Ref. \onlinecite{ahlrichs:98}, is to
vary this parameter in a set of simulations of a single monomer in
solvent (which can be done very easily), and to measure the momomer
diffusion coefficient $D_{0}$, until the latter has the desired value.
If viscosity and fluid density match as well, then the long--range
parts of the flow fields (beyond a few lattice spacings) must look the
same. It should be noted that the Einstein relation $D_{0} = k_{B} T /
\zeta_{\rm eff}$ thus defines an effective or renormalized friction
coefficient, which differs from the original bare one, as it contains
all the backflow effects. Since these tend to increase the mobility,
one has $\zeta_{\rm eff} < \zeta_{\rm bare}$. More quantitatively, one
can argue as follows: Let us consider a particle which is pulled
through the solvent at constant velocity $\vec{V}$ by an external
force $\vec{F}$.  Then, rewriting Eq. \ref{eq:ansatz}, we find
\begin{equation}
\vec{V} = \frac{1}{\zeta_{\rm bare}} \vec{F} + \vec{u}_{\rm av},
\end{equation}
where $\vec{u}_{\rm av}$ is the flow velocity averaged over the
nearest lattice sites of the particle, as implemented by our
interpolation procedure. However, to a good approximation, the flow
field should be given by the Oseen tensor:
\begin{equation}
\vec{u} = \frac{1}{8 \pi \eta r} \left( {\bf 1} + \vec{\hat{r}}
\otimes \vec{\hat{r}} \right) \vec{F},
\end{equation}
where $r$ is the distance from the particle. Hence the averaged flow
field should --- in our case of
averaging roughly at a distance $a$ from the particle --- have the form
\begin{equation}
\vec{u}_{\rm av} = \frac{1}{g \eta a} \vec{F} ,
\end{equation}
where $g$ is an unknown numeric constant describing the details of the
lattice geometry and of the averaging procedure. For example, doing
the average over a sphere of radius $d$, one would directly obtain
$\vec{u}_{\rm av} = \vec{F} / (6 \pi \eta d)$, from which one
easily derives Stokes' law. Combining these results and
using $\zeta_{\rm eff}\vec{V}=\vec{F}$ one obtains
\begin{equation} \label{eq:relationbareeff}
\frac{1}{\zeta_{\rm eff}} = \frac{1}{\zeta_{\rm bare}}
 + \frac{1}{g \eta a},
\end{equation}
i.~e. the overall mobility is simply the sum of the bare mobility and
a hydrodynamic, Stokes--type contribution, where the lattice
discretization serves to provide an effective Stokes radius of the
monomers. This relation has been tested by running several simulations
at different bare couplings and different lattice constants; the
agreement is remarkable, as seen from Fig. \ref{fig:zetaeffzetabare},
where we plot $\eta a / \zeta_{\rm eff}$ as a function of $\eta a /
\zeta_{\rm bare}$. The parameter $g$ is thus found to have the value
$g \approx 25$ for our method.

The lattice constant $a$ hence appears not only as a parameter which
controls how accurately the Navier--Stokes equation is solved (this is
the usual case for Navier--Stokes equation solvers), but it is being
assigned an additional meaning as an effective Stokes radius. For that
reason, it cannot be varied arbitrarily, but only within limits: A too
small lattice constant would result in an unphysically large particle
mobility, even if $\zeta_{\rm bare}$ is very large. This is quite
different from conventional Navier--Stokes equation solving, where one
obtains systematically better results when $a$ is decreased, and can
be viewed as the price which has to be paid for introducing the simple
and computationally fast concept of a point particle, which is
however, strictly spoken, unphysical. It should be noted that
$\zeta_{\rm bare}$ controls the degree of coupling to the flow field:
For small $\zeta_{\rm bare}$, one has $\zeta_{\rm eff} \approx
\zeta_{\rm bare}$, while for large $\zeta_{\rm bare}$ the Stokes
contribution prevails, $\zeta_{\rm eff} \approx g \eta a$. It should
have become clear that hence $\zeta_{\rm bare}$ has no real physical
meaning whatsoever; it is really the effective friction which matters
for the coupling.

\section{Single Chain Simulation}
\label{sec:singlechain}

\subsection{Input Parameters}
\label{sec:howto}

The present model is intended to represent the same physical situation
as an existing pure MD simulation \cite{duenweg:93}. We therefore
choose the physical input values for the new method as obtained by the
former (all values are given in the unit system specified in Sec.
\ref{sec:bead.spring}). The fluid is characterized by the the
temperature $k_B T = 1.2$, the density $\rho=0.864$, and the kinematic
viscosity $\nu=2.8$. The parameter $\mu$ (the fluid particle mass) is
unimportant; its value can be absorbed in a re--definition of the
$n_{i}$. The lattice constant $a$ of the grid is set to unity; this is
roughly the same as the bond length of the polymer chain, and the
interparticle distance of the MD fluid. As in the pure MD simulation,
we study chains of length $N_{\rm ch} = 30, 40$ and $60$. The
corresponding grid sizes (which are important parameters, since they
determine the hydrodynamic interaction of the chain with its periodic
images, see Ref. \onlinecite{duenweg:93}) are $L = 18$, $18$, and
$22$, respectively, which is roughly identical to the corresponding MD
box sizes.

The parameters for the FENE potential are taken from the MD simulation
as $R_0 = 2.0$ and $k = 7.0$. As already discussed in Sec.
\ref{sec:coupling}, this does however not assure that the static
conformations are identical: In the MD case, there is also the
influence of the solvent, which is absent in the present method.
Actually, the data show a systematic deviation, which is however not
very large (see Sec. \ref{sec:results.statics}).

The mass of the monomers was set to unity. This actually differs from
the MD case where the monomer mass had been set to two. However, we
also used a monomer of mass one in order to determine the ``bare''
friction coefficient $\zeta_{\text{bare}}$, using the procedure
outlined at the end of Sec. \ref{sec:coupling}, such that we found
$\zeta_{\text{bare}} = 20.8$ from the requirement that the monomer
diffusion coefficient has the value known from MD, $D_{0} = 0.076$.
Had we used a monomer of different mass, we would also have obtained a
slightly different value for $\zeta_{\rm bare}$ (these are very small
effects, beyond what the simple picture which underlies Eq.
\ref{eq:relationbareeff} can capture). Since however on the time scale
of Brownian motion it is only the parameter $\zeta_{\rm eff} = k_{B} T
/ D_{0}$ which matters, we expect an influence of the mass parameter
only for short times, where the dynamics differs from MD behavior
anyways.

It remains to specify the time steps $\Delta t$ and $\tau$. A choice
of $\Delta t = 0.01$ is optimal for the MD part \cite{kremer:90}.
Concerning the LBM time step $\tau$ it is desirable to make it as
large as possible because the fluid calculation is the CPU intensive
part of the method. Test simulations showed the limiting factor to be
that $n_i$ is getting negative for too large time steps due to
increasing fluctuations, in particular near the monomers. This
situation, however, can always happen, although with decreasing
probability for smaller time steps. We found that using a time step of
$\tau = 0.05$ only approximately each $10^4$th random number one $n_i$
became negative, while for $\tau = 0.01$ such a case never occurred
during the observation time. We decided to generate new random numbers
in such rare cases, which of course slightly changes the distribution
of the simulated noise, but is justified if the probability for
negative $n_i$'s is low enough. We ran the simulations at $\tau =
0.05$ and also did a simulation for the smallest system ($N_{\rm
ch}=30$, $L=18$) using $\tau = 0.01$ in order to check the results.

Furthermore, we should comment in some more detail on the lattice
constant $a$. The choice $a = 1$ seems intuitively reasonable, since
this matches the bond length and the interparticle distance in the MD
system. However, one would in principle like to make the lattice
spacing as large as possible, since, for constant overall volume, the
computational effort scales as $a^{-3}$. For this reason, we also did
a test run with $a = 2$ for the $N_{\rm ch} = 30$ system, where we of
course had re--adjusted the bare friction, see end of Sec.
\ref{sec:coupling}. It turned out that the decay of the dynamic
structure factor looks quite similar. However, there are systematic
discrepancies (see Sec. \ref{sec:comparisonofresults}), such that the
gain in speed is paid for by a certain loss in accuracy. In what
follows we will always refer to the case $a = 1$, unless explicitly
stated otherwise.
\end{multicols}
\widetext
\begin{table}[b]
\begin{tabular}{ccccc}
Chain length & 30 & 30 & 40 & 60 \\
LB time step $\tau$ & $0.05$ & $0.01$ & $0.05$ & $0.05$ \\
\tableline
exponent $\nu$ & $0.621 \pm 0.004$ & $0.620 \pm 0.002$ & $0.637 \pm 0.002$ & $0.637 \pm 0.002$ \\
$\left\langle R_{e}^{2}\right\rangle$ & $94 \pm 5$ & $90 \pm 4$ & $134 \pm 4$ & $217\pm 10$ \\
$\left\langle R_{g}^{2}\right\rangle$ & $14.3 \pm 0.5$ & $13.9 \pm 0.4$ & $20.6 \pm 0.3 $ & $33.5 \pm 0.9$ \\
$\left\langle\frac{1}{R_H}\right\rangle_{\infty}$ & $0.299 \pm 0.005$ & $0.300 \pm  0.005$ & $0.261 \pm 0.005$ & $ 0.215 \pm 0.004 $ \\
$\left\langle\frac{1}{R_H}\right\rangle_L$\tablenotemark[1] &
$0.1512$ & $0.1525$ & $0.1179$ & $0.0986$ \\
$k_{B}T$ & $1.139 \pm 0.003 $ & $1.2056 \pm 0.003$ & $1.139 \pm 0.003$ & $1.139 \pm 0.003$\\
$g_3$-exp.\tablenotemark[2] & $0.9951 \pm 0.0004$ & $1.009 \pm 0.0002$ &  $1.0001 \pm 0.0001$ &
$1.006 \pm 0.003$ \\ 
$g_1$-exp.\tablenotemark[2] & $0.6415 \pm 0.001$ & $0.6747 \pm 0.001$ & $0.6630 \pm 0.0006 $ & $0.6704 \pm 0.002$  \\
$D_{\text{CM}}$ & $6.533 \times 10^{-3} \pm 1 \times 10^{-5}$ & $6.102\times 10^{-3} \pm
1\times 10^{-5}$ & $4.860\times 10^{-3} \pm 2\times 10^{-5}$ & $3.387\times
10^{-3} \pm 1\times 10^{-5}$ \\
$D_{0}\tablenotemark[3] $ & $0.081$ & $0.062$ & $0.076$ & $0.054$\\
$\tau_Z$ (estimate) & $365$ & $380$ & $705$ & $1650$ \\
\end{tabular}
\tablenotetext[1]{no error due to complicated
  calculation}
\tablenotetext[2]{exponent obtained by fitting a power law in the
  sub-diffusive scaling regime $t\in[20:80]$}
\tablenotetext[3]{calculated using Eq. \ref{eq:fs.diffusion}}
\caption{Single chain properties}
\label{table:results}
\end{table}
\begin{multicols}{2}
\narrowtext
An important point concerning the comparison with analytical theory
should be mentioned here. It is usually assumed in these theories that
the time scale for the evolution of the hydrodynamic interaction is
much smaller than the diffusion time scale of a monomer, i.~e. the
Schmidt number $Sc=\frac{\nu}{D_{0}}\gg 1$. This parameter can be set
arbitrarily in our method: $\nu$ is an input parameter and $D_{0}$ can
be tuned by choosing $\zeta_{\text{bare}}$. In our case, we have $Sc
\approx 32$.

\subsection{Chain Statics}
\label{sec:results.statics}

The results for the chain lengths of $N_{\rm ch} = 30$, $40$ and $60$
are listed in Table \ref{table:results}. The measurement of the
chain's temperature provides a first consistency check of the
algorithm. The values for $k_B T_{\text{measured}} \equiv
\frac{2}{3 N_{\text{ch}}} E_{\text{kin}}$ show a discretization error
of 5\% for the large time step $\tau = 0.05$. For the small time step
$\tau = 0.01$ the error decreases significantly.

\begin{figure}
\begin{center}
\epsfig{file=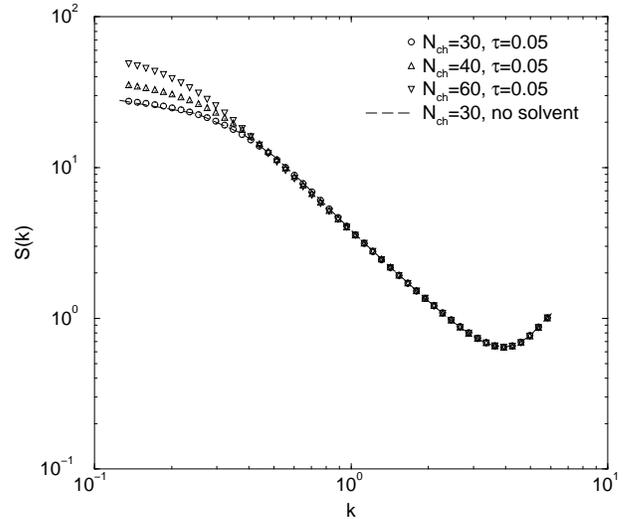,width=7cm,angle=-90}
\end{center}
\caption{The static structure factor of the chains.}
\label{fig:static.strfac}
\end{figure}

The radius of gyration
\begin{equation}
\left< R_G^2 \right> = \frac{1}{2 N_{\rm ch}^2} \sum_{ij}
\left< r_{ij}^2 \right>,
\end{equation}
with $r_{ij} = \left\vert \vec{r}_i - \vec{r}_j \right\vert$, and the
end--to--end distance
\begin{equation}
\left\langle R^2_{e}\right\rangle =
\left\langle\left(\vec{r}_{N_{\rm ch}}-\vec{r}_{1}\right)^2\right\rangle
\end{equation}
are related to the number of monomers by the static exponent $\nu$,
\begin{equation} \label{eq:gyration.scaling}
\left\langle R_{g}^{2}\right\rangle \propto 
\left\langle R^2_{e}\right\rangle \propto
N_{\rm ch}^{2 \nu} ;
\end{equation}
for a self--avoiding walk $\nu \approx 0.588$ from renormalization
group theory methods and Monte Carlo simulation \cite{sokal:95}. In
principle, $\nu$ can be obtained from the scaling law
(\ref{eq:gyration.scaling}); however, this would require simulations
covering a wide range of $N_{\rm ch}$. Hence, it is advantageous to
use the static structure factor
\begin{eqnarray}
S(k) &=& N_{\rm ch}^{-1}\sum_{ij}
\left\langle\exp(i\vec{k}\cdot\vec{r}_{ij})\right\rangle \\ \nonumber
     &=& N_{\rm ch}^{-1}\sum_{ij}
\left\langle\frac{\sin({k}{r_{ij}})}{{k}{r_{ij}}}\right\rangle,
\end{eqnarray}
which probes different length scales even for a single polymer. In
the scaling regime $R_{g}^{-1} \ll k \ll a_{0}^{-1}$ ($a_{0}$ being a
microscopic length of the order of the bond length) the relation
\begin{equation}
S(k) \propto k^{-1/\nu}
\end{equation}
holds \cite{doi:86}. By fitting a power law to our data (see
Fig. \ref{fig:static.strfac}) we get the values for $\nu$ of Table
\ref{table:results} which are about 6\% higher than the asymptotically
correct value, resulting from the finite chain length. In
Fig.~\ref{fig:static.strfac} we also include data which have been
generated from a simulation of a single chain {\em without}
surrounding LBM fluid. The conformations must be the same, i.~e. the
structure factors must coincide (up to discretization errors, which
may look somewhat different for the chain coupled to the LBM
fluid). As is seen from the figure, the agreement is very good,
i.~e. the method is validated to produce correct static conformations.

The hydrodynamic radius 
\begin{equation}
\left\langle\frac{1}{R_{H}}\right\rangle_{\infty} = 
\frac{1}{N_{\rm ch}^{2}}\sum_{i\neq j}
\left\langle\frac{1}{r_{ij}}\right\rangle
\end{equation} 
is an interesting quantity because the Kirkwood prediction for the diffusion
of the chain's center of mass \cite{kirkwood:48,kirkwood:58}
\begin{equation}\label{eq:kirkwood.formula}
D_{\text{CM}}=
\frac{D_{0}}{N_{\rm ch}}+\frac{k_{B}T}{6\pi\eta}
\left\langle\frac{1}{R_{H}}\right\rangle_{\infty}
\end{equation} 
depends on it. This formula, however, is only correct for a single
chain in an infinite medium. In a finite box one has to take into
account the hydrodynamic interaction with the periodic images. This
will eventually lead to a finite--size corrected hydrodynamic
radius. Quite generally, one must expect a finite size effect of order
$L^{-1}$ for every dynamic quantity, corresponding to the slow
$r^{-1}$ decay of hydrodynamic interactions. A detailed description
can be found in Refs. \onlinecite{duenweg:93,beenakker:86}, so that we
can restrict ourselves to the essential points. Within the Oseen
approximation, the diffusion tensor is given by
\begin{equation}\label{eq:diffusion.matrix.1}
{\bf D}_{ij} \equiv {\bf D}(\vec{r}_{ij})= 
\frac{k_{B}T}{\eta L^{3}}\sum_{\vec{k}\neq 0}
\frac{{\bf 1}-\vec{\hat{k}}\otimes\vec{\hat{k}}}{k^{2}}
\exp(i\vec{k}\cdot\vec{r}_{ij})
\end{equation}
for $i \neq j$, where $\vec{k} = 2 \pi \vec{n} / L$ ($\vec{n}$ being a
vector of integers) runs over the reciprocal lattice vectors and
$\vec{\hat{k}}$ is a unit vector in the direction of $\vec{k}$. For $i
= j$, one has the monomeric diffusion coefficient $D_0$, plus the
contribution due to the hydrodynamic interaction of that bead with its
own periodic images,
\begin{equation}\label{eq:diffusion.matrix.2}
{\bf D}_{ii} =
D_{0}{\bf 1}+\lim_{\vec{r}\rightarrow0}
\left({\bf D}(\vec{r})-\frac{k_{B}T}{8 \pi \eta r}
({\bf 1}+\vec{\hat{r}}\otimes\vec{\hat{r}})\right) .
\end{equation}
The last two expressions can be calculated efficiently using the Ewald
summation technique. The center of mass diffusion constant is given by
\begin{equation}\label{eq:cm.diffusion}
D_{\text{CM}} = \frac{1}{N_{\rm ch}^{2}}\sum_{ij}
\frac{1}{3}{\text{Tr}}\left\langle {\bf D}_{ij}\right\rangle.
\end{equation}
Inserting Eqs. \ref{eq:diffusion.matrix.1} and \ref{eq:diffusion.matrix.2}
one obtains \cite{duenweg:93}
\begin{equation}\label{eq:fs.diffusion}
D_{\text{CM},L}=\frac{D_{0}}{N_{\rm ch}}
-\frac{2.837\,k_{B}T}{6\pi\eta L N_{\rm ch}}
+\frac{1}{3N_{\rm ch^{2}}}\sum_{i\neq j}
{\text{Tr}}\left\langle {\bf D}_{ij}\right\rangle ,
\end{equation}
which defines, by comparison with the Kirkwood formula
(\ref{eq:kirkwood.formula}), a finite size corrected hydrodynamic
radius:
\begin{equation}
D_{\text{CM},L} \equiv \frac{D_{0}}{N_{\rm ch}}
+\frac{k_{B}T}{6\pi\eta} \left\langle\frac{1}{R_{H}}\right\rangle_{L}.
\end{equation}
$R_H$ is thus effectively increased by the periodic images. For our
box sizes, the discrepancy between $\left\langle R_{H}^{-1}
\right\rangle_{L}$ and $\left\langle R_{H}^{-1}
\right\rangle_{\infty}$ amounts to approximately a factor of two
(cf. Table \ref{table:results}). This is in agreement with the
corrections found in Ref. \onlinecite{duenweg:93}.
\begin{figure}
\begin{center}
\epsfig{file=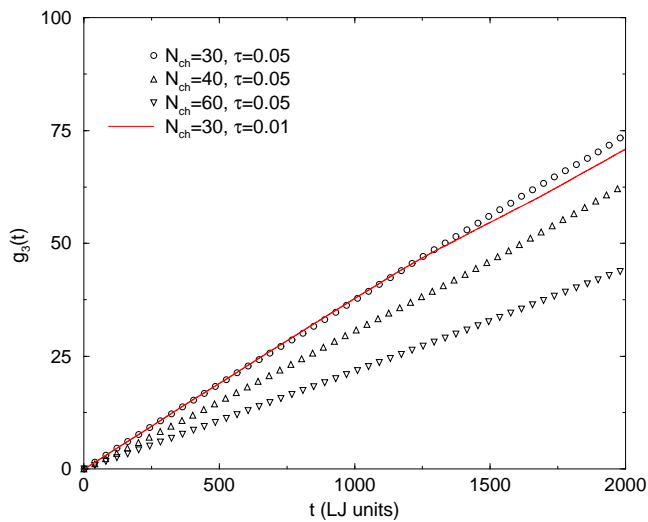,width=7cm,angle=-90}
\end{center}
\caption{The mean square displacement of the chain's center of mass.}
\label{fig:cm.diffusion}
\end{figure}

\subsection{Chain Dynamics}

The dynamic scaling picture for Zimm dynamics \cite{doi:86} starts
from the prediction $D_{\rm CM} \propto R_g^{-1}$ (cf. Eq.
\ref{eq:kirkwood.formula}). The Zimm time $\tau_Z$, i.~e. the longest
relaxation time of the chain, is given by the condition that the chain
has moved its own size during $\tau_Z$, or $D_{\rm CM} \tau_Z \propto
R_g^2$, implying $\tau_Z \propto R_g^3$, which defines the dynamic
exponent $z = 3$. This exponent then quite generally relates times to
corresponding lengths, such that, for example, the mean square
displacement of a monomer on time scales below $\tau_Z$, but above the
microscopic time scales $\tau_0$, should be proportional to $t^{2/z} =
t^{2/3}$. For a chain without hydrodynamic interaction (Rouse model),
where $D_{\rm CM} \propto N_{\rm ch}^{-1}$, one finds $z = 2 + 1 /
\nu$ from analogous considerations.

Figure \ref{fig:cm.diffusion} shows the mean square displacement of
the chain's center of mass
\begin{equation}
g_3(t) = \left\langle
\left(\vec{R}_{\text{CM}}(t_0+t)-\vec{R}_{\text{CM}}(t_0)\right)^2
\right\rangle.
\end{equation}
By fitting a power law we obtain the exponents and the diffusion
constants shown in Table \ref{table:results}. Obviously, the exponents
support the prediction of simple diffusive behavior ($t^1$). One would
expect theoretically that two diffusive regimes exist, both exhibiting
$t^1$ behavior but different prefactors, with a smooth crossover
around the Zimm time. The accuracy of the data does not allow to
support this crossover, which is not surprising as the short--time and
long--time diffusion constant are expected to be rather close to each
other \cite{rey:91,fixman:81,fixman:83}. In principle, the scaling
behavior of $D_{\text{CM}}$ provides a test of the Zimm prediction
$D_{\text{CM}} \propto N_{\rm ch}^{-\nu}$. But there are large
corrections to scaling due to finite chain length and bead size
effects \cite{duenweg:93,kremer:89}. Therefore it is more useful to
analyze the non--asymptotic relation (\ref{eq:fs.diffusion}) by
comparing the values for $D_{0}$ that can be obtained from Eq.
\ref{eq:fs.diffusion}, where finite chain length and finite box size
are taken into account, with the input value of $D_{0}=0.076$. The
values are also listed in Table \ref{table:results} showing quite
reasonable agreement. Without the finite size corrections, the
agreement is unacceptable, such that a negative value for $D_0$ would
be obtained.

\begin{figure}
\begin{center}
\epsfig{file=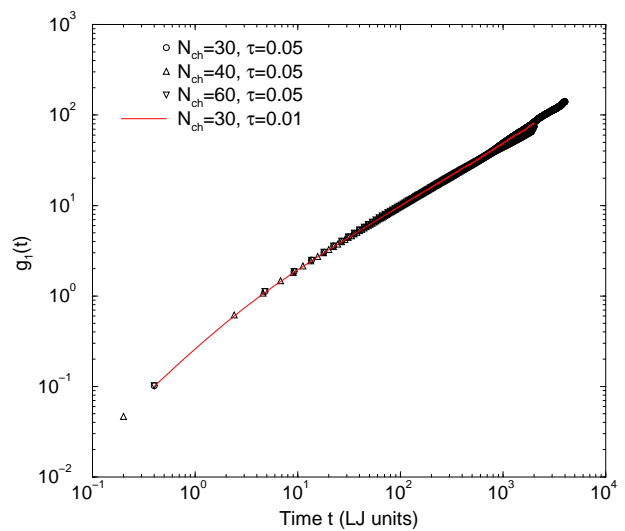,width=7cm,angle=-90}
\end{center}
\caption{The mean square displacement of the central monomer.}
\label{fig:middle.diffusion}
\end{figure}

The mean square dis\-place\-ment of a single mo\-no\-mer $i$ (which
should only be evaluated for monomers near the center of the chain to
eliminate end effects)
\begin{equation}
g_1(t) = \left\langle
\left(\vec{r}_i(t+t_0)-\vec{r}_i(t_0)\right)^2\right\rangle
\end{equation}
is plotted in Fig. \ref{fig:middle.diffusion}. In the time regime
below the Zimm time and above the ballistic regime, the scaling
behavior $g_1(t) \propto t^{2/3}$ is predicted. The corresponding fit
to our data yields the exponents of Table \ref{table:results}. The
values obviously favor the Zimm model compared to the Rouse model,
which predicts $g_1(t) \propto t^{2/z} = t^{0.54}$.

\begin{figure}
\begin{center}
\epsfig{file=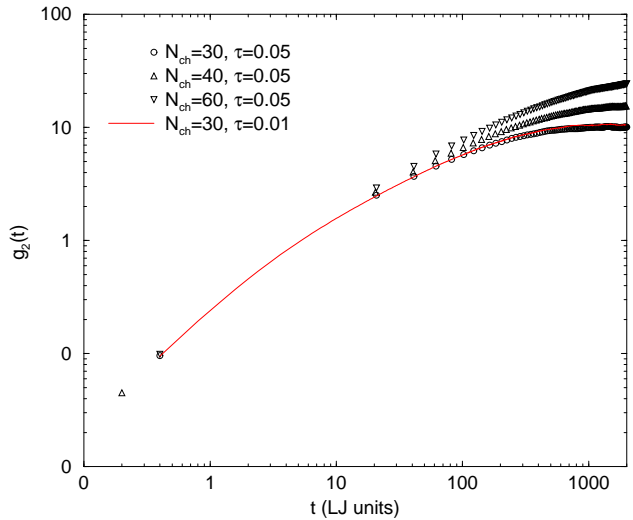,width=7cm,angle=-90}
\end{center}
\caption{The mean square displacement of the central monomer in
  the chain's center of mass system.}
\label{fig:middle.cm.diffusion}
\end{figure}

The Zimm time can be estimated from the mean square displacement of
a monomer in the center of mass system,
\begin{eqnarray}
g_2(t) =
\left\langle\right. && \left(
  \left[
    \vec{r}_i(t+t_0)-\vec{R}_{\text{CM}}(t+t_0) \right] \right.\\\nonumber
  &&\left.\left.-\left[ \vec{r}_i(t_0)-\vec{R}_{\text{CM}}(t_0) \right]
    \right)^2 \right\rangle ,
\end{eqnarray}
which is depicted in Fig. \ref{fig:middle.cm.diffusion}.
Theoretically, a crossover to a plateau should evolve at the Zimm
time. However, the crossover is quite extended in our simulation,
making it difficult to extract a specific time for it. We therefore
estimate the Zimm time from
\begin{equation}
\tau_Z = \frac{\left\langle R_g^2\right\rangle}{6 D_{\text{CM}}} ,
\end{equation}
which yields the values shown in Table \ref{table:results}.

It is interesting to perform a Rouse mode analysis. For this purpose one
defines the Rouse modes as \cite{kopf:97}
\begin{equation}\label{eq:def.rouse.modes}
\vec{X}_p = N_{\rm ch}^{-1} \sum_{n=1}^{N_{\rm ch}}\vec{r}_n \cos\left[
  \frac{p\pi}{N_{\rm ch}}(n-\frac{1}{2})\right] .
\end{equation}
It is well known that these modes are the (independent) eigenmodes of
the random walk Rouse model \cite{doi:86}.
\begin{figure}\vspace*{0.2cm}
\begin{center}
\epsfig{file=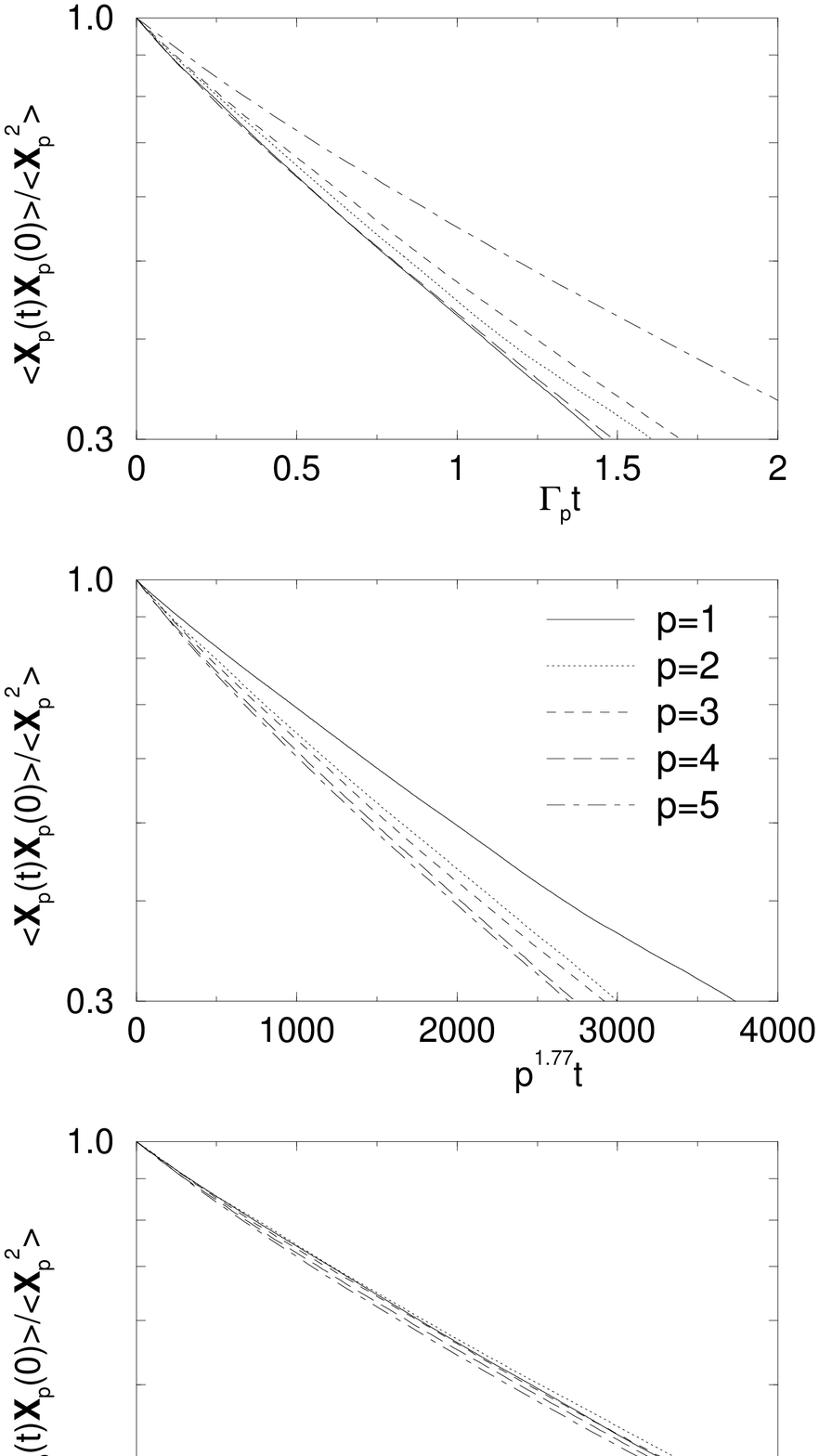,width=7cm}
\end{center}\vspace*{2cm}
\caption{Normalized autocorrelation function of the Rouse mode
  $\vec{X}_p$ for different $p$, for the longest simulated chain
  $N_{\rm ch} = 60$. The upper part of the figure uses $\Gamma_p t$
  as scaling argument, where $\Gamma_p$ was calculated directly from
  the chain conformations. The middle part uses $p^{z \nu} t$, where
  naive dynamic scaling has been applied, while the lower part also
  takes the correction factor $r(p)$ (see Appendix
  \ref{app:initdecay}) into account. The meaning of symbols is the
  same for all three parts (see middle part).}
\label{fig:rouse.modes}
\end{figure}
However, for reasons of
translational symmetry along the chain, one must expect that the cross
correlation
$\left\langle\vec{X}_p(t+t_0)\vec{X}_q(t_0)\right\rangle\,(p \neq q)$
is at any rate quite weak, regardless of chain statistics and
dynamics, such that the modes can be viewed as independent modes even
beyond the random walk Rouse case. For a ring polymer, this can be
shown rigorously, since in this case there is strict invariance under
the transformation $n \to n + 1$, such that the Rouse modes (which are
then defined with an $\exp\left(i p \pi n / N_{\rm ch} \right)$
factor) are eigenfunctions under this transformation. Hence, if end
effects are not too strong, one should also expect for our case an
independence of the Rouse modes. Indeed, within the accuracy of our
data, the cross correlation terms are zero.

Furthermore, within the approximations of the Zimm model, the
autocorrelation function of the modes should decay exponentially
\cite{doi:86},
\begin{equation}
\frac{\left\langle\vec{X}_p(t+t_0)\vec{X}_p(t_0)\right\rangle}
{\left\langle\vec{X}_p^2\right\rangle}
  = \exp(-t/\tau_p) .
\end{equation}
In Fig. \ref{fig:rouse.modes}, we therefore plot, for $p \ge 1$, the
normalized autocorrelation function semi--logarithmically as a function
of properly scaled time. Firstly, we estimate $\tau_p$ via the initial
decay rate
\begin{equation}
\tau_p^{-1} = \Gamma_p = -\frac{d}{dt}\left.\left(
\frac{\left\langle \vec{X}_p(t)\vec{X}_p(0) \right\rangle}
     {\left\langle \vec{X}_p^2              \right\rangle}
     \right)\right |_{t=0} ,
\end{equation}
which can, within the framework of Kirkwood--Zimm theory, be
calculated in terms of purely static averages, i.~e. from the chain
conformations in combination with a model diffusion tensor, for which
we use Eqs.~\ref{eq:diffusion.matrix.1} and
\ref{eq:diffusion.matrix.2}. The details of this approach are
described in Appendix~\ref{app:initdecay}. Interestingly, it turns out
that this quantity is only subject to an $L^{-3}$ finite size effect
(which we neglect), in contrast to the usual $L^{-1}$ behavior. This
result holds beyond the various approximations of
Appendix~\ref{app:initdecay}; our interpretation is that any
contribution of global center--of--mass motion of the chain is being
subtracted, such that the leading--order hydrodynamic interaction with
the periodic images cancels out, and only a dipole--type interaction
remains. In the upper part of Fig.~\ref{fig:rouse.modes}, we thus plot
the autocorrelation as a function of $\Gamma_p t$, where $\Gamma_p$
was calculated directly from the simulated chain conformations, in
combination with the Oseen tensor. It is seen that the Oseen formula
describes the decay quite well; however, the data collapse is not
particularly good. There is also some curvature, indicating a
non--exponential decay. The middle part of the figure then uses the
scaling argument $p^{z \nu} t$. This $p$--dependence results from the
calculation of $\Gamma_p$, where instead of the actual chain
conformations asymptotic self--avoiding walk statistics is employed
(see Appendix~\ref{app:initdecay}), as the leading power law. This
corresponds to simple dynamic scaling, which views the $p$th mode as
equivalent to a chain of length $N_{\rm ch} / p$, such that $\tau_p
\propto (N_{\rm ch} / p)^{\nu z}$. However, the more detailed
calculation of Appendix~\ref{app:initdecay} yields an additional weak
$p$--dependence, i.~e. a correction factor $r(p)$, whose presence
indicates, in our opinion, that the simple picture of subchains of
length $N_{\rm ch} / p$ is not fully justified. Taking this correction
into account, we obtain a very nice data collapse (see lower part of
Fig.~\ref{fig:rouse.modes}). This is quite remarkable; one would of
course expect the best data collapse for the uppermost part which
involves the smallest number of approximations. It seems that there
are various errors involved which somehow happen to cancel out.
\begin{figure}\vspace*{0.2cm}
\begin{center}
\epsfig{file=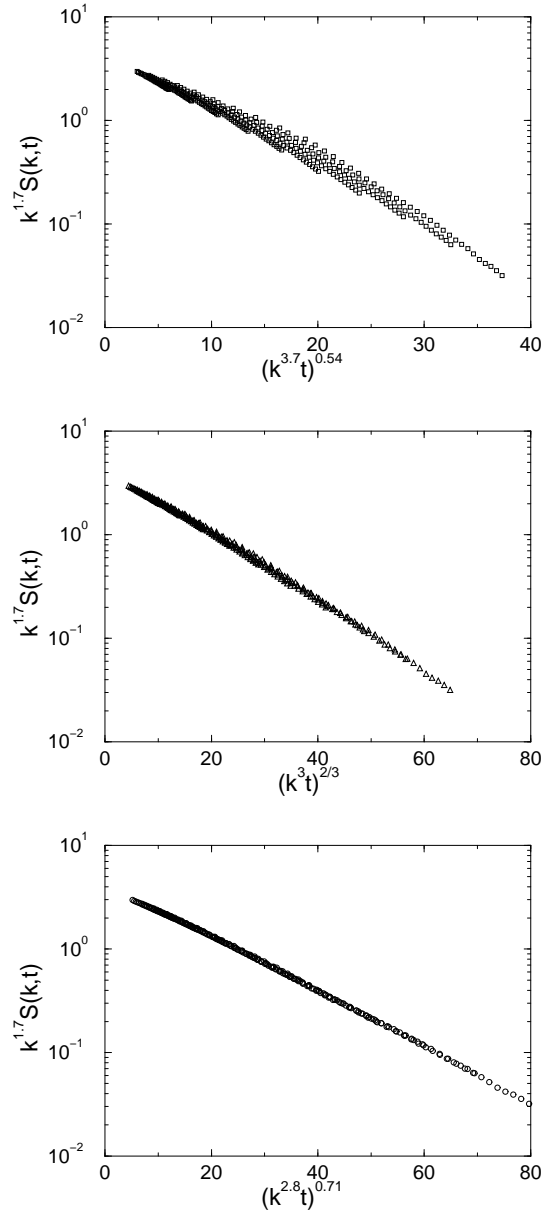,width=7cm}
\end{center}\vspace*{2cm}
\caption{Scaling plot of the dynamic structure factor for
  $N_{\rm ch} = 60$, for Rouse scaling ($z = 3.7$), asymptotic
  Zimm scaling ($z = 3$), and $z = 2.8$, which produces the best
  collapse.}
\label{fig:zimm.line.collapse.n60}
\end{figure}
As far
as the absolute value of the decay rate is concerned we find
reasonable agreement: While the lower part of
Fig.~\ref{fig:rouse.modes} shows a decay rate of roughly $3 \times
10^{-4} p^{3 \nu} r(p)$, Eq.~\ref{eq:gammapfinalresult} predicts a
decay rate of order $5.4 \times 10^{-4} p^{3 \nu} r(p)$ where we have
for simplicity used the random walk value for the constant $A$, and
$b^3=2.0$ (extracted from the results for $R_e^2$ via $R_e^2=b^2
N^{2\nu}$).

The dynamic structure factor
\begin{equation}
S(k,t) = \frac{1}{N_{\rm ch}}\sum_{ij}\left\langle \exp\left(i
  \vec{k}\cdot\left[\vec{r}_{i}(t)-\vec{r}_{j}(0)\right]\right)
  \right\rangle
\end{equation}
is predicted \cite{doi:86} to exhibit the scaling behavior
\begin{equation}
S(k,t) = S(k,0) f(k^{z}t)
\end{equation}
if both wavenumber and time are in the scaling regime, i.~e.
$R_g^{-1} \ll k \ll a_0^{-1}$ and $\tau_0 \ll t \ll \tau_Z$. It is
even possible to calculate explicit formulas (rigorously for the
random walk Rouse model and using the linearization approximation in
the Zimm case) \cite{doi:86,degennes:67,akcasu:80}, which suggest that
there is an exponential dependency on $(k^{z}t)^{2/z}$ for $\Gamma_k t
\gg 1$, where $\Gamma_k$ is the ($k$--dependent) decay rate. Hence a
plot of $S(k,t) k^{1/\nu}$ against $(k^{z}t)^{2/z}$ should --- for the
correct model --- collapse to a straight line in a log--linear
representation. For $N_{\rm ch} = 60$, the results are shown in Fig.
\ref{fig:zimm.line.collapse.n60} (the plots for the other chain lengths
look quite similar). The data were restricted to the scaling regime
$20 \leq t \leq 80$ and $0.7 \leq k \leq 2$. These ranges were
obtained from the single--monomer mean square displacement, Fig.
\ref{fig:middle.diffusion}, and from the static structure factor, Fig.
\ref{fig:static.strfac}, respectively. Values of $S(k,t)$ below $0.01$ were
discarded, for reasons of statistical accuracy. It is clearly visible
that the simulation shows Zimm rather than Rouse behavior. A dynamic
exponent of $z = 2.8$ yields the best data collapse. Such an effective
value, which is, due to corrections to scaling, somewhat smaller than
the correct asymptotic one, is quite usually observed, not only in
simulations \cite{pierleoni:92}, but also in experiments \cite{degennes:79}.

Concerning finite size effects, one has for a finite box size
$S = S(k,t,L)$, and scaling is corrupted by the second length $L$ in the
problem. The influence can be estimated, in close analogy to the
procedure presented in Appendix \ref{app:initdecay}, by studying
the Akcasu formula for the $k$--dependent diffusion coefficient
\cite{akcasu:80,benmouna:80},
\begin{equation}
D(k,L) = \frac{\sum_{ij}\left\langle \vec{\hat{k}}\cdot {\bf D}_{ij}\cdot 
\vec{\hat{k}} \exp(i\vec{k}{\vec{r}_{ij}})\right\rangle}
  {\sum_{ij}\left\langle\exp(i\vec{k}{\vec{r}_{ij}})\right\rangle},
\end{equation}
which is $L$--dependent because of the finite size form
(\ref{eq:diffusion.matrix.1}) of ${\bf D_{ij}}$. $D(k,L)$ is related to the
initial slope of the dynamic structure factor via
\begin{equation}
D(k,L) = -\lim_{t\rightarrow 0}\frac{1}{k^2 t}
\ln\left(\frac{S(k,t,L)}{S(k,0,L)}\right)
\end{equation}
We do not present the details of our semi--quantitative analysis here
since they have been outlined in Ref. \onlinecite{duenweg:93} already.
The result is a $k$--independent correction term of order $L^{-1}$
(note that $S(k,t)$ does contain the overall chain motion, for every
wavenumber). As the leading--order ($L = \infty$) term is proportional
to $k$ in the scaling regime, the conclusion is that scaling is
corrupted, but the relative contribution of the finite size correction
gets weaker with increasing $k$. For the $k \rightarrow 0$ limit
finite size corrections amount to roughly $100\%$, as has been shown
by the calculations for the hydrodynamic radius in Sec.
\ref{sec:results.statics}. In the scaling regime the corrections are
much smaller, because it is closer to the $k L \rightarrow \infty$
limit. This is, in our opinion, the main reason why the data collapse
works so nicely.

\section{Comparison to the Corresponding MD System}
\label{sec:mapping}

\subsection{Efficiency} \label{sec:efficiency}

Since the system is highly dilute, the CPU cost for the MD part for
the polymer chain is negligible, and the lattice Boltzmann part uses
up practically all computational resources. It should be noted that
this part can be optimized by choosing appropriate simulation
parameters; our choice ($a = 1$, $\tau = 0.05$) is probably not the
most efficient one. Firstly, it is possible to increase the lattice
spacing somewhat, without substantial loss in accuracy. For example,
going from $a = 1$ to $a = 2$ reduces the computational effort by a
factor of eight. This increase seems however to be slightly too large
already; as outlined in Sec. \ref{sec:comparisonofresults}, $a = 2$
produces less accurate data. Secondly, one can try to exploit
Eq.~\ref{eq:relationnulambda} by varying $\tau$ or $a$ while keeping
$\nu=2.8$, such that the simulation runs at $\lambda=-1$, for which the
LBM algorithm takes a particularly simple form in which a substantial
number of operations can be saved\cite{ladd:94}. Further speedup can be
expected if the requirement $\nu=2.8$ and $D_0=0.076$ (for mapping to
MD) were released. However,
we have not checked these questions in a systematic fashion; in
particular, our discussion has not taken into account that the limit
of stable time steps $\tau$ depends on both $a$ and $\lambda$ in a
non--trivial way. We hence want to simply state that our present
choice of parameters is not yet a fully optimized one; therefore the
numbers given below (for $a = 1$ and $\tau = 0.05$) should be viewed
as a lower bound of the efficiency which the method can attain.

On one EV5.6 processor of a 433~MHz DEC Alpha server 8400 (for a
typical box size of $L = 40$) our code obtains $3.1 \times 10^5$ grid
point updates per second. In order to compare this number with the
molecular dynamics system, we note that one grid point corresponds to
$0.86$ solvent particles for $\rho = 0.86$ and $a = 1.0$. Therefore,
the efficiency of the code in MD units is $3.1 \times 10^5 \times 0.86
\approx 2.7 \times 10^5$ particle updates per second. This number
should be contrasted with the efficiency of optimized MD codes for
short--range LJ fluids, which is \cite{soddemann:98} (on the same
machine) $2.1 \times 10^5$ particle updates per second, using the code
described in Ref. \onlinecite{puetz:98}. Thus, the LBM would run by a
factor of $1.3$ faster than MD if the same time step were used.
However, the lattice Boltzmann time step $\tau = 0.05$ is more than an
order of magnitude larger than for the pure MD system: The latter must
be run without friction and noise, i.~e. in the microcanonical
ensemble, in order to strictly conserve momentum (otherwise the
hydrodynamic interaction would be screened \cite{duenweg2:93}). Such a
simulation can only be stable on long time scales if the time step is
sufficiently small; according to our experiences \cite{kopf:97}, one
needs $\Delta t = 0.003$. Taking these factors into account, we obtain
a net speedup of a factor of $22$, which, as outlined above, can be
increased further by choosing a coarser lattice, i.~e. by trading in
accuracy for speed. A detailed comparison with the ``competitor'' DPD
\cite{hoogerbrugge:92,schlijper:95,warren:97,espanol:95,marsh:97,%
  espanol:98,ignacio:98} is highly desirable, but not done here, last
not least because the match of the viscosity is much less trivial in
DPD \cite{marsh:97,espanol:98}. From what we know from the literature,
we expect that the two methods would be roughly comparable in speed,
at least by order of magnitude.

\begin{figure}
\begin{center}
\epsfig{file=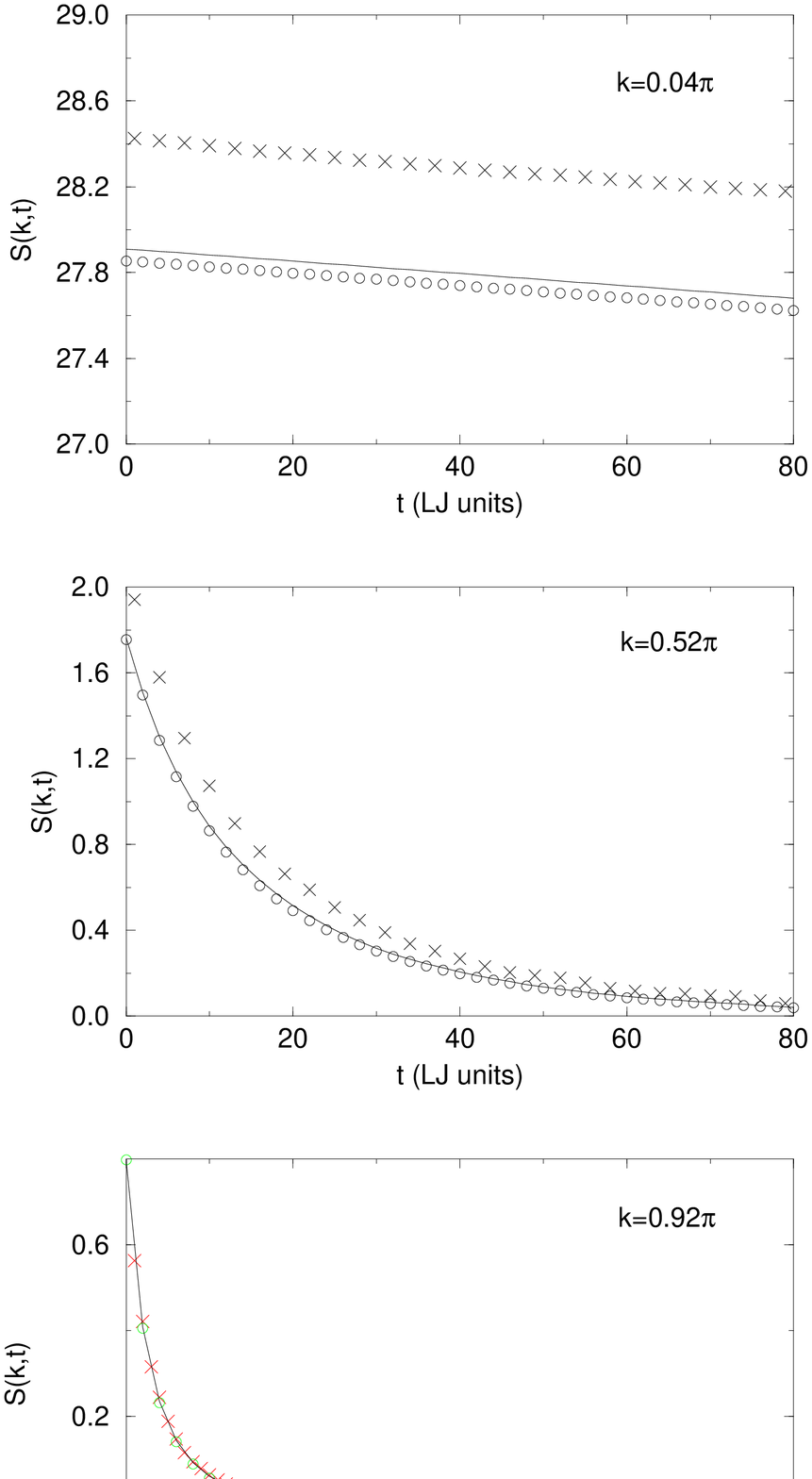,width=7cm}
\end{center}\vspace*{2cm}
\caption{The dynamic structure factor $S(k,t)$ for the new 
  method with $\tau=0.05$ (circles) and $\tau=0.01$ (line) compared to
  pure MD simulation (crosses) for three different $k$ values ($N_{\rm
  ch}=30$).}
\label{fig:mapping.strfct.k}
\end{figure}

\subsection{Static and Dynamic Behavior}
\label{sec:comparisonofresults}

\begin{figure}\vspace*{0.2cm}
\begin{center}
\epsfig{file=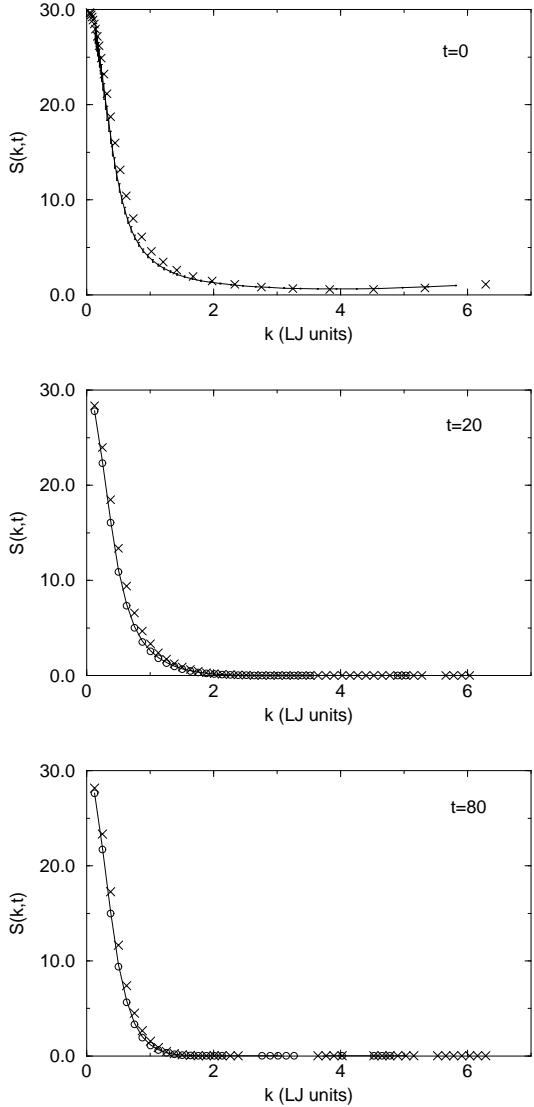,width=7cm}
\end{center}\vspace*{2cm}
\caption{The dynamic structure factor $S(k,t)$ for the new method
  with $\tau=0.05$ (circles) and $\tau=0.01$ (line) compared to pure
  MD simulation (crosses) for three different times ($N_{\rm ch}=30$).}
\label{fig:mapping.strfct.t}
\end{figure}

\begin{figure}
\begin{center}
\epsfig{file=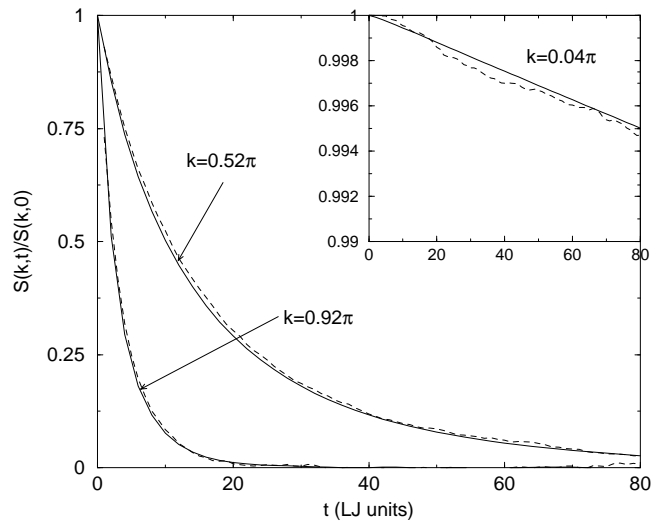,width=7cm,angle=-90}
\end{center}
\caption{$S(k,t) / S(k,0)$ for $N_{\rm ch} = 60$ using the new method
  (solid lines) with $\tau=0.05$ compared to pure MD simulation
  (dashed lines) for three different $k$ values.}
\label{fig:mapping.strfct.w/o.statics}
\end{figure}

In order to check how well the new method produces the same physics as
the original MD model \cite{duenweg:93}, from which all simulation
parameters were derived, we focus on the comparison of the structure
factor $S(k,t)$ for both methods, as shown in Fig.
\ref{fig:mapping.strfct.k} (time dependence at constant $k$),
Fig. \ref{fig:mapping.strfct.t} ($k$ dependence at constant time),
and Fig. \ref{fig:mapping.strfct.w/o.statics} (time dependence
for the {\em normalized} structure factor).

Let us first consider the {\em static} case $t = 0$. The corresponding
plot (Fig. \ref{fig:mapping.strfct.t}, uppermost part) for $N_{\rm ch}
= 30$ shows systematic deviations. These manifest for example in the
discrepancies of the static scaling exponent ($\nu=0.59$ for the pure
MD simulation, $\nu = 0.62 \ldots 0.64$ for the new method); the chain
is more stretched using the new method. The absolute values for the
static structure factor differ up to about $25\%$. Similar results
hold if one compares other static quantities like the radius of
gyration or the end--to--end distance. It can be verified that the
discrepancies show no significant dependency on the chain length for
the range investigated here ($30$--$60$). Moreover, they are not due
to a discretization error in time, as the plots for $\tau=0.05$ and
$\tau=0.01$ show. The reason is rather simply the fact that the MD
chain is subject to a different potential (intra--chain plus solvent)
than the LBM chain (intra--chain only). For that reason, there is a
systematic difference in the static conformations, which then, in
turn, will also affect the dynamic properties somewhat. For example,
the $N_{\rm ch} = 60$ chain has a gyration radius $\left< R_g^2
\right>^{1/2} = 5.79$, while the corresponding MD chain
\cite{duenweg:93} has a gyration radius of only $4.78$. It is hence
not surprising that the larger chain is also somewhat slower, as the
comparison of the diffusion constants confirms ($D_{CM} = 3.39 \times
10^{-3}$ for the larger LBM chain, and $D_{CM} = 4.25 \times 10^{-3}$
for the smaller MD chain). Therefore, in order to achieve a better
match of static and dynamic properties, it would be necessary to
re--adjust the potential for the LBM chain such that the conformations
are more similar. This is possible, but not completely trivial, and
has not been attempted in this work. On the other hand, for the
dynamics parameters (i.~e. the viscosity and the friction
coefficient), it is quite easy to achieve matching, as has been
described in Sec. \ref{sec:howto}.

Turning to the decay of $S(k,t)$, we first note that the direct
comparison of the data (Fig. \ref{fig:mapping.strfct.k} and
\ref{fig:mapping.strfct.t}) yields similar discrepancies of
up to 25\% as for the static case. The overall agreement is however
quite reasonable. In order to divide out the trivial amplitude effect,
we also plot $S(k,t)/S(k,0)$ for three different $k$ values in
Fig. \ref{fig:mapping.strfct.w/o.statics}. For $k$ in the scaling
regime, the agreement is much better, with differences of a few
percent only. This is not too surprising, since in this regime the
decay rate should in essence be given by $k^3 k_B T / \eta$ times a
numerical prefactor which depends only weakly on the details of the
chain statistics \cite{akcasu:80,benmouna:80}. In the long--wavelength
regime (inset of Fig. \ref{fig:mapping.strfct.w/o.statics}) the decay
is given by $\exp \left( - D_{\rm CM} k^2 t \right)$, which is
nicely confirmed by the data, and thus the ratio of the decay rates
is just the ratio of the diffusion constants, i.~e. there is again
a discrepancy of roughly 20 \% (this is hardly visible in Fig.
\ref{fig:mapping.strfct.w/o.statics}, due to noise in the MD data).

To summarize, we find that both methods are well--suited for
quantitatively reproducing the dynamics of polymer chains in solvent,
and both reveal Zimm behavior very nicely. The discrepancies which we
find in the dynamic properties can be directly traced back to the
non--perfect match of the static conformations. If those had been
matched by an adjustment of the potential, then the agreement would
probably be close to perfect.

\begin{figure}
\begin{center}
\epsfig{file=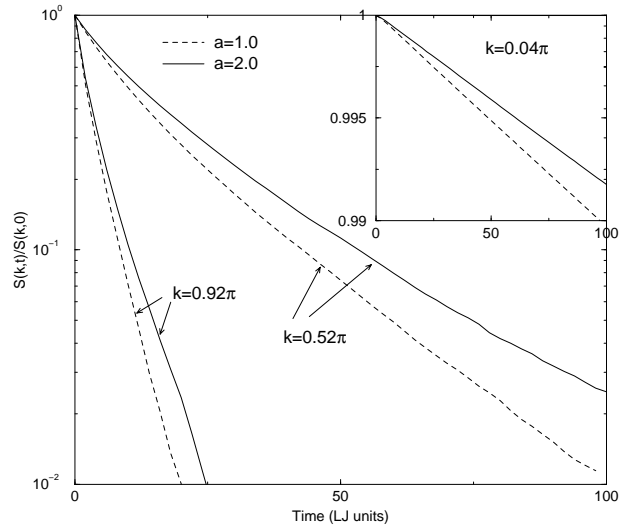,width=7cm,angle=-90}
\end{center} 
\caption{Comparison of $S(k,t) / S(k,0)$ for $N_{\rm ch} = 30$,
  two different lattice spacings, and three $k$ values as indicated.}
\label{fig:compa1a2}
\end{figure}

Finally, let us discuss in more quantitative terms the influence of
the lattice spacing. To this end, Fig. \ref{fig:compa1a2} compares the
decay of the normalized dynamic structure factor of an $N_{\rm ch} =
30$ chain for three $k$ values, obtained by running the same system
with two different lattice spacings $a = 1$ (as discussed previously)
and $a = 2$. All other simulation parameters (in particular the box
volume, and the monomeric diffusion coefficient $D_0$ --- {\em not}
the bare coupling $\zeta_{\rm bare}$) were left identical. As one sees
from the figure, the larger lattice spacing induces decays which are
systematically slower, by roughly 20 \% to 25\%. It is thus a question
of desired accuracy if one wants to consider these results as still
acceptable or not. The observed effect goes in the direction which one
expects, for the following reasons: As soon as the lattice spacing
exceeds the size of the chain, there will be no hydrodynamics left and
one will observe pure Rouse dynamics, which is slower. Of course, this
must be a systematic crossover as a function of lattice spacing. Thus
one expects a decrease of the hydrodynamic correlations with
increasing $a$ (also consistent with the reasoning at the end of Sec.
\ref{sec:coupling}), and hence a systematic slowdown of the dynamics.

\section{Conclusion and Outlook}
\label{sec:conclusion}

With this paper we have established a new method to simulate
polymer--solvent systems. The solvent is modeled by the lattice
Boltzmann method and the polymer by a continuum bead--spring model.
The two parts are coupled using a simple dissipative friction ansatz
which locally conserves mass and momentum. The driving force of the
system are thermal fluctuations which are added to both the fluid and
the polymer. The main advantage of the new method compared to MD is
its computational efficiency, which amounts to a factor of 20, or even
more, if one is willing to be satisfied with less accurate results.

As described in Sec. \ref{sec:howto}, it is possible to obtain the
physical input parameters for the new method from results of existing
MD simulations. Therefore, one can view the present method as a
coarse--graining procedure where one goes in a well--controlled way from
small length and time scales to larger ones. As the results show, this
is possible without substantial loss of information about the statics
and dynamics on the mesoscopic scale.

The input which is needed from a more microscopic approach consists
of: (i) Effective potentials for the coarse--grained monomers such
that the static chain conformations are roughly reproduced (this was
the part to which we did not pay much attention, with the result that
this is the largest source for the observed deviations); (ii) the
solvent temperature, density and viscosity, and (iii) the monomeric
diffusion coefficient, from which one adjusts the coupling.

It seems that a lattice spacing which roughly matches the chain's bond
length and the interparticle distance of the solvent is optimal. A
lattice constant which is chosen too large will result in
underestimated hydrodynamic interactions, as seen from the data with
$a = 2$, while a too small lattice spacing will result in a large
computational effort, plus (if it becomes very small) a monomeric
diffusion coefficient which will exceed any realistic value, due to an
effective Stokes radius which is too small.

We have chosen the parameters of Ref. \onlinecite{duenweg:93} for our
simulation and performed a detailed quantitative comparison of the
results. The main deviations result from insufficient match of the
static conformations. The current model is therefore as appropriate as
the original MD model for verification of Zimm dynamics in dilute
polymer solutions. The dynamic scaling laws (in particular the $k^3 t$
decay of the dynamic structure factor) could be observed, and there is
good agreement with the decay rates predicted by the Zimm model, if
the finite box size effects are taken into account. Interestingly
enough, the decay of the Rouse modes is only subject to an $L^{-3}$
finite size effect, while most other decay rates have a large $L^{-1}$
finite size correction, due to the $r^{-1}$ behavior of the Oseen
tensor.

After having tested the method successfully future work can now deal
with more controversial problems, like the influence of hydrodynamics
on the motion of a semi--flexible chain or the hydrodynamic screening
in semidliute solutions. It should however be kept in mind that the
algorithm in its current version is only suitable for problems where
the polymer concentration is low. The coupling only takes into account
the momentum transfer between monomers and solvent. Excluded--volume
effects between solvent particles and monomers, which are very
important for processes like, e.~g., the penetration of solvent into a
dense polymer matrix, are not properly modeled. A study of such topics
would require a generalization of the algorithm which would assign a
finite volume to the monomers.

It is a pleasure to thank Ralf Everaers and Alexander Kolb for helpful
discussions, and the latter for a critical reading of the manuscript.

\appendix

\section{Initial Decay Rate of Rouse Modes}
\label{app:initdecay}

In this appendix we outline the details of the calculation of $\Gamma_{p}$,
i.~e. the initial decay rate of the autocorrelation function of the Rouse
modes for $p \ge 1$, where we treat the general case of a chain whose
statistics is described by an exponent $\nu$ (i.~e. $\nu = 0.5$ for a random
walk (RW), and $\nu = 0.6$ for a self--avoiding walk (SAW)). We start by
stating the result of linear response theory \cite{doi:86},
\begin{eqnarray}
\Gamma_p & = &-\frac{d}{dt}\left.\left(
\frac{\left\langle \vec{X}_p(t)\vec{X}_p(0) \right\rangle}
     {\left\langle \vec{X}_p^2              \right\rangle}
     \right)\right |_{t=0}  \\ \nonumber
         & = & \frac{1}{\left\langle \vec{X}_p^2 \right\rangle}
     \sum_{i,j,\alpha,\beta,\gamma} \left\langle
     \frac{\partial {X}_{p\gamma}}{\partial {r}_{i\alpha}}
     D_{ij\alpha\beta}
     \frac{\partial {X}_{p\gamma}}{\partial {r}_{j\beta}} \right\rangle,
\end{eqnarray}
where Greek indices again denote Cartesian coordinates.
Evaluating the derivatives of the Rouse modes, one obtains
\begin{eqnarray} \label{eq:gammapwithtrace}
\Gamma_p = \frac{1}{\left< \vec{X}_p^2 \right> N_{\rm ch}^2} 
    \sum_{i,j} \cos\left(\frac{p\pi}{N_{\rm ch}}(i-1/2)\right)\\\nonumber
               \cos\left(\frac{p\pi}{N_{\rm ch}}(j-1/2)\right)
    {\rm Tr} \left< {\bf D}_{ij} \right> .
\end{eqnarray}
From the definition of the Rouse modes, Eq. \ref{eq:def.rouse.modes},
one finds
\begin{eqnarray}
\left< \vec{X}_{p}^{2} \right> = \frac{1}{N_{\rm ch}^{2}} \sum_{ij}
\left< \vec{r}_{i} \cdot \vec{r}_{j} \right>
\cos\left(\frac{p\pi}{N_{\rm ch}}(i-1/2)\right)\\\nonumber
\cos\left(\frac{p\pi}{N_{\rm ch}}(j-1/2)\right) ,
\end{eqnarray}
which is evaluated via ($b$ is the bond length)
\begin{eqnarray}
\vec{r}_{i} \cdot \vec{r}_{j} & = & \frac{1}{2} \left[
\vec{r}_{i}^{2} + \vec{r}_{j}^{2} - 
\left( \vec{r}_{i} - \vec{r}_{j} \right)^{2} \right]  \\
0 & = & \sum_{i=1}^{N_{\rm ch}} \cos\left(
   \frac{p\pi}{N_{\rm ch}}(i-1/2) \right) \label{eq:sumofcosines} \\
\left< \left( \vec{r}_{i} - \vec{r}_{j} \right)^{2} \right> & = &
b^{2} \left\vert i - j \right\vert^{2 \nu}
\end{eqnarray}
(note that the last relation holds only asymptotically for large
$\left\vert i - j \right\vert$). Approximation by an integral yields
\begin{eqnarray}
\left< \vec{X}_{p}^{2} \right> = - \frac{b^{2}}{2 N_{\rm ch}^{2}}
\int_{0}^{N_{\rm ch}} dx \int_{0}^{N_{\rm ch}} dy
\left\vert x - y \right\vert^{2 \nu}\\\nonumber
\cos\left(\frac{p\pi}{N_{\rm ch}} x \right)
\cos\left(\frac{p\pi}{N_{\rm ch}} y \right) .
\end{eqnarray}
Furthermore, we use the relation
\begin{equation}
\cos \alpha \cos \beta = \frac{1}{2} \left[ 
\cos \left( \alpha - \beta \right) + 
\cos \left( \alpha + \beta \right) \right]
\end{equation}
and transform to the variables
\begin{equation}
u = \frac{p \pi}{N_{\rm ch}} (x - y), \hspace{1cm}
v = \frac{p \pi}{N_{\rm ch}} (x + y) .
\end{equation}
Exploiting the symmetry of the integrand with respect to $u$, and
performing the integration over $v$, we find
\begin{equation}
\left< \vec{X}_{p}^{2} \right> =
\frac{b^{2} N_{\rm ch}^{2\nu}}{2 (p \pi)^{1 + 2\nu}} f(p)
\end{equation}
with
\begin{equation}
f(p) = \frac{1}{p \pi} \int_{0}^{p \pi} du\, u^{2 \nu}
\left[ \sin u - (p \pi - u) \cos u \right] .
\end{equation}
For the RW case, $f(p)$ is exactly unity, while for the SAW case
a weak dependence on $p$ remains; however, also in this case $f(p)$
is close to one. Using the MAPLE software package, we have
numerically evaluated this function; for the first 20 Rouse
modes it is tabulated in Table \ref{table:corrfactorstable}.

The calculation of the numerator of Eq. \ref{eq:gammapwithtrace} is
performed using precisely the same procedure, the only difference
being that $\left< \left( \vec{r}_{i} - \vec{r}_{j} \right)^{2}
\right>$ is replaced by ${\rm Tr} \left< {\bf D}_{ij} \right>$, which
we calculate using the finite box size form, Eq. \ref{eq:diffusion.matrix.1}:
\begin{equation}
{\rm Tr} \left< {\bf D}_{ij} \right> =
\frac{k_{B} T}{\eta \pi^{2}} \int_{k_{0}}^{\infty} dk
\left< \exp ( i \vec{k} \cdot \vec{r}_{ij} ) \right> ,
\end{equation}
where we have replaced the summation over wavenumbers by an integral
\begin{equation}
\frac{1}{L^{3}} \sum_{\vec{k}\neq 0} \rightarrow 
\frac{1}{(2\pi)^3} \int_{k_0}^{\infty} 4 \pi k^2 dk, 
\end{equation}
$k_{0} = 2 \pi / L$ denoting the cutoff wavenumber due to the finite
box size.

The factor $\left< \exp ( i \vec{k} \cdot \vec{r}_{ij} ) \right>$
describes the structure of the chain, and must, for reasons of
scaling \cite{degennes:79}, asymptotically have the form
\begin{equation}
\left< \exp ( i \vec{k} \cdot \vec{r}_{ij} ) \right> = 
g\left( k^{2} b^{2} \left\vert i - j \right\vert^{2 \nu} \right) .
\end{equation}
It should be noted that, for reasons of inflection symmetry, $g$ must depend
on $k^{2}$, and that $g(0) = 1$. We further introduce the constants
\begin{eqnarray}
A & = & \int_{0}^{\infty} dw g(w^{2}) \\
B & = & \left. \frac{d g(w^2)}{d w^2} \right\vert_{w = 0} .
\end{eqnarray}
For example, for a random walk one has
$g = \exp \left( - (b^{2} / 6) k^{2} \left\vert i - j \right\vert \right)$,
i.~e. $g(w^{2}) = \exp( - w^{2} / 6)$, $A = \sqrt{3 \pi / 2}$,
$B = - 1 / 6$. We now calculate ${\rm Tr} \left< {\bf D}_{ij} \right>$
by performing a Taylor expansion with respect to $k_{0} = O(L^{-1})$; the
result is
\begin{eqnarray}\nonumber
{\rm Tr} \left< {\bf D}_{ij} \right> &=& 
\frac{k_{B} T}{\eta \pi^{2}} \left[
\frac{A}{b \left\vert i - j \right\vert^{\nu}} - k_{0}
- \frac{B}{3} b^{2} \left\vert i - j \right\vert^{2 \nu} k_{0}^{3}\right]\\
&& + O(k_{0}^{5}).
\end{eqnarray}
Interestingly, the linear term does not depend on the monomer indices
at all. From this, we conclude that the linear $L^{-1}$ contribution
to the decay rate exactly vanishes, due to Eq. \ref{eq:sumofcosines},
and that the leading order finite size effect is actually of order
$L^{-3}$, i.~e. quite small. In what follows we will therefore only
concentrate on the leading--order term for an infinite box.
Using the same procedure as for $\left< \vec{X}_p^{2} \right>$,
one finds
\begin{table}
\begin{tabular}{c|c|ccc} 
    & RW            & \multicolumn{3}{c}{SAW}        \\ 
$p$ & $h(p) = r(p)$ & $f(p)$   & $h(p)$   & $r(p)$   \\ \hline
 1  & 1.040901      & 1.229939 & 1.531335 & 1.245049 \\
 2  & 1.155368      & 1.096321 & 1.671897 & 1.525007 \\
 3  & 1.186325      & 1.099453 & 1.711021 & 1.556248 \\
 4  & 1.203640      & 1.075431 & 1.732468 & 1.610952 \\
 5  & 1.213328      & 1.077224 & 1.744639 & 1.619569 \\
 6  & 1.220118      & 1.067140 & 1.753074 & 1.642778 \\
 7  & 1.224789      & 1.068286 & 1.758929 & 1.646496 \\
 8  & 1.228399      & 1.062691 & 1.763420 & 1.659391 \\
 9  & 1.231138      & 1.063494 & 1.766850 & 1.661363 \\
10  & 1.233376      & 1.059915 & 1.769636 & 1.669601 \\
11  & 1.235174      & 1.060514 & 1.771886 & 1.670781 \\
12  & 1.236696      & 1.058018 & 1.773782 & 1.676515 \\
13  & 1.237967      & 1.058484 & 1.775371 & 1.677278 \\
14  & 1.239069      & 1.056638 & 1.776745 & 1.681508 \\
15  & 1.240014      & 1.057013 & 1.777926 & 1.682029 \\
16  & 1.240849      & 1.055589 & 1.778967 & 1.685283 \\
17  & 1.241579      & 1.055899 & 1.779880 & 1.685654 \\
18  & 1.242234      & 1.054765 & 1.780696 & 1.688239 \\
19  & 1.242815      & 1.055026 & 1.781422 & 1.688510 \\
20  & 1.243341      & 1.054101 & 1.782079 & 1.690615 \\ 
\end{tabular}
\caption{The functions $f(p)$, $h(p)$, and $r(p)$, as defined
  in the text, for both the RW and the SAW case.}
\label{table:corrfactorstable}
\end{table}
\begin{eqnarray}\nonumber
\int_0^{N_{\rm ch}} dx \int_0^{N_{\rm ch}} dy && 
\frac{1}{\left\vert x - y \right\vert^\nu}
\cos\left(\frac{p\pi}{N_{\rm ch}} x \right)
\cos\left(\frac{p\pi}{N_{\rm ch}} y \right) =\\
&& = \frac{N_{\rm ch}^{2 - \nu}}{(p \pi)^{1 - \nu}} h(p)
\end{eqnarray}
with
\begin{equation}
h(p) = \frac{1}{p \pi} \int_0^{p \pi} du
\frac{1}{u^\nu} \left[ (p \pi - u) \cos u - \sin u \right] .
\end{equation}
This function also exhibits a weak $p$--dependence, see Table
\ref{table:corrfactorstable}, even for a RW. Finally, introducing
\begin{equation}
r(p) = h(p) / f(p),
\end{equation}
also tabulated in Table \ref{table:corrfactorstable},
we can write the result for $\Gamma_p$ as
\begin{equation} \label{eq:gammapfinalresult}
\Gamma_p = A \frac{2}{\pi^2} \frac{k_B T}{\eta b^3}
\left( \frac{p \pi}{N_{\rm ch}} \right)^{3 \nu} r(p) .
\end{equation}
The leading power--law dependence on $p$ and $N_{\rm ch}$ is exactly
what one expects from dynamic scaling. The function $r(p)$ is a
correction to scaling. As far as the numerical prefactor is concerned,
we get (in the RW case) a relaxation which is roughly the same as that
calculated in the textbook by Doi and Edwards~\cite{doi:86}.

\end{multicols}

\end{document}